\documentstyle[12pt,aasms]{article}
\tighten

\def\rmeq#1{\eqno({\rm #1})}
 
\def\lc{light curve}
\def\ml{microlensing} \def\mo{monitoring} \def\pr{program} \def\mlmpr{\ml\ \mo\ \pr} 
\def\ev{event} 
\def\ex{expansion}
  
\def\bi{binary}  \def\bis{binaries} 
\def\rd{Di\thinspace Stefano} 
\def\bl{binary lens} 

\def\ct{coefficient}
\def\cc{caustic crossing}
\def\mag{magnification}
\def\pm{point mass}

\begin{document}
\title{Identifying Microlensing by Binaries  
}
\author{Rosanne Di\thinspace Stefano and Rosalba Perna}
\affil{Harvard-Smithsonian Center for Astrophysics, Cambridge, MA 02138}

\begin{abstract}
The \mlmpr s have studied large numbers of standard \lc s which seem to be due
to lensing by a dark point mass. 
Theory predicts that many \ml\ \ev s should display significant
deviations from the standard form. Lens binarity in particular is expected to 
be common.
So far, however, only a handful of \lc s exhibit
evidence that the lens is a binary; all of these display dramatic deviations
from the standard \lc , exhibiting pronounced multiple peaks and/or
caustic crossings. 

Binary-lens \ev s in which the \lc\ 
is less dramatically
perturbed should also exist in the data set. Why, then, have we not
detected them? 
The answer may lie in the 
fact that the perturbations, though often significant,
tend to be
less distinctive than those associated with caustic crossings. 
It is therefore possible that some of these more gently perturbed \ev s
have been misclassified, and that others have simply been missed.
Reliable estimates of the overall detection efficiency, hence derivation
of the fraction of the Halo mass that may be in MACHOs, rely on
resolving this issue.  Microlensing can also be used to determine the form
of the initial mass function (IMF) beyond the solar neighborhood; accurate
determination of the IMF also relies on the ability to correctly identify
binary light curves.

We present a method to determine whether a \lc\ 
is due to lensing by a binary. The method works for both gently
and dramatically perturbed \bl\ \lc s. 
Our method identifies all degenerate solutions--i.e., all possible
lensing \ev s that might have given rise to the observed \lc. It also
enables us to eliminate from consideration large ranges of possible false
positive identifications associated with \lc s that might mimic \ml\
by a \bi .
This method, or a generalization of it, can also be applied to the analysis of
\lc s that deviate from the standard point-mass lens form because of
astronomical effects other than lens binarity. 
\end{abstract}

\section{Introduction}

\vspace{-.2 cm}

Given a light curve, how can we determine whether it is associated
with a microlensing event? It is relatively easy to answer this question 
if a single dark mass lenses a point source in an otherwise
blank field (see, e.g., Paczy\'nski 1986). 
This, however, is not the case expected to be most commonly
encountered in the microlensing observations. Source and lens binarity
(Griest \& Hu 1992; Mao \& Paczy\'nski 1991;  Mao \& \rd\ 1995;
Udalski {\it et al.} 1994; \rd\ \& Mao 1996; \rd\ 1996),
blending (Udalski {\it et al.} 1994; \rd\ \& Esin 1995; Kamionkowski 1995;
Buchalter, Kamionkowski, \& Rich 1996), and finite source size effects
(Witt \& Mao 1994; Witt 1995) are expected to play a role in
a large fraction of microlensing events. The associated light curves
will not be of the standard form. Some will be relatively mild
perturbations of the standard form. Others will be dramatically
different.  
The presence of these non-standard \lc s in the data sets presents a
challenge to the analysis of the ensemble of events.  Without systematic
methods to reliably recognize that they do correspond to microlensing
events, the detection efficiency is diminished and its 
numerical value uncertain.
If the observing teams are to achieve their goal of quantifying the size and
characteristics of compact dark objects in the Local Group, it is therefore
crucial that they develop the ability to discover and study a wide range of
non-standard light curves in addition to those due to lensing by a point
mass.

In this paper we focus on the problem of identifying \lc s
due to lensing by a \bi . A first attempt to solve this problem was made
by Mao \& \rd\ (1995), who found that the timing of maxima and minima
could be used
as a first step toward identifying all possible physical events that might have
given rise to a multiply-peaked \lc , particularly if some of the peaks
are  due to caustic crossings. Their method was used to study the first
detected binary lens (Udalski {\it et al.} 1994; Alcock {\it et al.} 1996). 
Today, the observing teams each have the
capability of analyzing  strongly perturbed binary-lens \lc s. 
Until now, however, the more gently
perturbed \lc s which should also be present in the data set have neither been
classified nor detected. 

\subsection{Expected Numbers}

Figure 1 shows 15 light curves. These have been randomly
generated, with all of the physical variables (\S 2.1)
chosen from
uniform probability distributions. (See the figure caption.) 
The \lc s with sharply rising and falling wall-like structures 
represent \ev s in which the track of the source has crossed through
caustic curves. It is interesting that OGLE 7, the first 
discovered \bl\ \ml\ \ev\ (Udalski {\it et al.} 1994; 
Alcock {\it et al.} 1996) exhibits the
double wall structure found in 3 of the \lc s shown.
It is also interesting that approximately $8$ of these
$15$ randomly chosen curves are  relatively gentle
perturbations of the standard \pm\ lens \lc , of a sort that
has not yet been reliably identified in the data set. 
The fractions expected in the data sets of single-peak and multiple-peak \lc s,
of \lc s exhibiting \cc s, and of \lc s 
exhibiting the more gentle perturbations
cannot be inferred from this graph.  
The number of \lc s shown is too small, and the underlying
distribution of values of mass ratios and orbital separations among
the population of \bi\ lenses is not likely to be uniform. 
Nevertheless, this simple exercise 
provides insight consistent with a more detailed
consideration of the statistics of binary lensing (\rd\ 1997).   
We also note here that observational studies indicate that, for
non-mass-transfer binaries containing low-mass
stars,
the
distribution of values of $q$ may peak toward small values
(e.g., $q^{-1}$ [Trimble 1990], or a Gaussian centered on
$q=0.23$ [Duquennoy \& Mayor 1991]).  
If such distributions apply to a significant number of binary
lenses, then 
the fraction of
gently perturbed \lc s becomes larger.
The reason for this is that the linear dimensions of typical
caustic structures become smaller as $q$ decreases.

This brief discussion makes it clear that it is important for the
observing teams to have at their disposal a method of 
reliably identifying and interpreting {\it all} binary-lens \lc s. 
 
\subsection{Plan of the Paper}

Section~2 describes our classification scheme and its
application. 
Given a \lc , our goal is to find any and all binary-lens \lc s that
provide acceptable fits to it. 
Smooth \lc s are considered in \S 2.3  and 
extremely perturbed \lc s, including those with \cc s,
are considered in \S 2.4.

 We find  that the \lc s
are highly degenerate, in that many distinct physical lensing \ev s
can create essentially identical \lc s. An important feature 
of our method is that
it identifies all physical solutions that are equivalent up to any chosen 
level as a function of photometric precision.   
In \S 3 we address the general question of distinguishing between true
microlensing \lc s and others that may have similar characteristics.  
We find that it is possible to use our
method to eliminate many ``pretenders'' as well as to identify the possible
physical parameters of true binary-lens events.
In \S 4 we discuss the implications and possible extensions of this work.

\section{Classification Scheme}

\vspace{-.2 cm}

\subsection{Physical Characterization of the Lensing Events} 
 
\vspace{-.2 cm}

The physical event that gives rise to a \lc\ can be characterized
by the following quantities: (1) the binary's total mass, 
$M=m_1+m_2$ (with $m_1 > m_2$); 
(2) the ratio of the components'
mass, $q=m_2/m_1$; (3) $a_{\rm orb}$, 
the projected separation on the lens plane between
$m_1$ and $m_2$; (4) the distance of closest approach, $b$, between the track
of the source and the binary's center of mass; and (5) the angle of approach, 
$\theta$, as measured with respect to the binary axis (with the direction 
from $m_1$ to $m_2$ defining $\theta =0$). We will normalize to $M=1$, thus
allowing us to specify any lensing \ev\ by the set of physical
parameters   
${p}=\{q, a_{\rm orb}, b, \theta\}$, 
where the first two elements describe properties
directly related to the binary and its orientation in space, and the last
two describe characteristics of the lensing event. All distances are measured
in terms of $R_E$, the Einstein radius associated with the total mass, $M$.  
In principle, the full orbital solution of the \bi\ serving
as the lens could come into play if, for example, the time duration of the \ev\
is a considerable fraction of the orbital period.  
We return to this point in \S 4. 

\subsection{Classification of the Light Curves}

\vspace{-.2 cm}

Our thesis is
that, if a \ml\ model fits the data, then there is a high probability that the
\lc\ is actually due to \ml .  
The viability of this conjecture is discussed in \S 3. Use of the
conjecture implies, however, that the primary challenge is to identify all 
\ml\ \lc s that fit the data. A simple method to do this
is introduced in \S 2.3. This straightforward approach is all
that is needed to identify all \lc s that fit any smooth
function. 
The application of an extended approach to
extremely perturbed \lc s, including those exhibiting singularities, is described in \S 2.4.

\subsubsection{A Note on Degeneracy}

In general, genuine binary \ml\ \lc s can be fit by
many different lens models; as for the point-mass lens, there is a high
degree of degeneracy.  Also as in the case of the point lens, the degeneracy
is due to symmetry. The symmetry is of two types: first is symmetry
inherent in the pattern of isomagnification contours.  For example, several
angles of approach may be equivalent slices through the geometry of a given
lens.  Secondly, there are slices through the lens plane of different lenses
that (because of scaling, for example) are essentially equivalent.  Because
of this degeneracy, it is not adequate to find one or even a few possible
binary models.  Instead, we would like to find all binary lens models; this
not only ensures that the correct model is among the ones we know about,
it also allows the use of probability methods to study properties of the lens
population.

\subsubsection{Classification}

The set of physical parameters, $\{p_i\}$, describes an event,
$i$, that gave rise to
a binary-lens \lc . The \lc\ itself can be characterized by
a separate set of parameters, which describe its
shape, independent of the physical event that produced it. 
For the sake of clarity, we will refer to this second set
of parameters as the \lc\ parameters. For example, the coefficients
of a polynomial expansion provide a complete set of \lc\ parameters
for smooth \lc s.  

The basic idea behind our method is to characterize an observed \lc\
by a set of \lc\ parameters 
and to identify all \bl\
 \lc s characterized, within the limits of observational
uncertainty, by the same parameters.

\subsection{Expansion of A Smooth Light Curve}  
\vspace{-.2 cm}

Our approach is mathematical, but 
can be applied to real \lc s.
The basic idea is to treat each \lc\ as a mathematical
function $A(t)$ which can be expanded in terms of a complete set of
orthonormal functions
$$
A(t)=\sum^\infty_{n=0}a_nT_n(t)\;,\rmeq{1a}
$$
where,
$$
\int^\infty_{-\infty} T_n(t)T_j(t) w(t)\, dt=\delta_{nj}\;, \rmeq{1b}
$$
$w(t)$ is a weight function whose form is specific
to the set of base functions ${T_n(t)}$,
and
$$a_n=\int^\infty_{-\infty} A(t) T_n(t) w(t)\, dt.\rmeq{1c}$$    
In this way, every \lc\ is completely specified by the values of the set of
coefficients $\{ a_n\}$.  We may thus view the \lc s as residing in the
space of the coefficients.

This mathematical expansion not only provides a systematic way to
characterize each light curve, but also to quantify the degree of similarity 
between different \lc s. 
If two light curves are similar (i.e., of similar shape and encompassing
nearly equal areas)
then 
the coefficients of their expansions
will be close in a way that can be quantified. Indeed, the condition
$$\int^\infty_{-\infty}\Big[A(t)-A^\prime(t)\Big]^2 w(t)\, dt < \epsilon
\rmeq{2a}$$ 
is equivalent to
$$D^2=\sum^\infty_{n=0}\Big[a_n-a^\prime_n\Big]^2 < \epsilon.
\rmeq{2b}$$  
Thus, distances in the space of the \lc s can be measured with a
Euclidean metric. (Bolatto \& Falco, e.g., used an observation-based approach  
parallel to this.)

While any complete, orthonormal set of functions can be used to expand
the \lc s, some choices have advantages over others. We have chosen
to use the Tchebyshev polynomials, ${T_n(x)}$. 
\footnote{The Tchebyshev polynomials can be expressed as
$T_n(x)=cos\,(n\, arccos(x))$. They are defined and orthogonal on the
interval [-1,1], over a weight $w(x)=(1-x^2)^{-1/2}$.
We have $T_0(x)=1$, $T_1(x)=x$; the 
recursion relation $[T_{n+1}(x)=2\, x\, T_n(x) -T_{n-1}(x)]$ 
can be used to derive the functional form of $T_n$ for $n>1$.
$T_n$ has $n$ zeros and $n+1$ extrema on the interval $[-1,1]$.
}
The special charm of the Tchebyshev polynomials for the
expansion of smooth \lc s is that the convergence is rapid, and,
in addition, for smooth \lc s the expansion can be truncated in such a way
as to mimic photometric errors of a predetermined size, consistent
with the observations of any given \lc . This is as follows.

Suppose that we have expanded a light curve, which has been frequently sampled, 
to order $N$. 
It is important to
know how well the \lc\ can be approximated if we truncate the expansion
at lower order, $m$. 
{Indeed, since the $T_k(x)$'s are all bounded between plus and minus
$1$, the 
discrepancy can be no larger than the sum of the 
absolute value of the neglected $a_k$'s
(k=m+1,....N). If the $a_k$'s are rapidly decreasing (which is the 
case for smooth \lc s, see also Fig. 2), then the error is dominated 
by $a_{m+1}T_{m+1}(x)$,
an oscillatory function with m+2 equal extrema distributed smoothly
over the interval
[-1,1]. Thus, the error is spread uniformly over the whole interval.
We can now use this formalism to mimic the observational uncertainties.
Given the size of photometric errors (say $\delta=1\%$), we can terminate the
expansion of the \lc\ at the order at which $a_{m+1} \sim \delta$ (where $a_0$
has been normalized to unity). A similar cut-off should also be applied to
those 
numerically-generated \lc s to which the observed curve is being compared, so that they can,
in effect, be sampled with a photometric precision comparable to that
achieved for the data.

Given an observed \lc , we compute a Tchebyshev \ex\
as follows. 
First, 
the time interval over  which the light curve is sampled must be mapped
onto the interval [-1,1]. This can be done by studying the ``wings" of the \lc,
as the magnification just begins to rise above (and then to fall below)
unity; we have chosen the cut-off magnification to be $1.06$,
corresponding to a distance of $2 R_E$ from a point mass lens.
The interval $[t_{up}, t_{down}]$ (from the baseline just
before the
\ev\ begins to the baseline just after the \ev\ ends) is mapped onto
the interval [-1,1].
Then, we compute the \ex\ \ct s, ${a_n}$. Instead of using the expression 
in (1a), we do a somewhat different \ex: 
$$A(t)=\sum^m_{n=0}a_nT_n(t) -{{1}\over{2}}a_0.\rmeq{3}$$     
This alternative
expansion has the useful feature that it is guaranteed to
exactly agree with the observed \lc\ at $m$ points, the positions of
the $m$ zeros of $T_m$ (see, e.g.,
Press {\it et al.} 1992).  
 
\vspace {-.5 cm}
 
\subsubsection{Equivalent Light Curves}
\vspace {-.3 cm}
The $\{a_n\}$ characterize the \lc, to whatever level of precision we
have chosen. Our ultimate goal is to use this characterization to identify
all \bl\ \lc s which, to the same level of precision, are equivalent
to the original one. If the \lc\ we started with, drawn from the data,
is indeed due to lensing by a \bi , then we would like to find all
sets of the physical parameters $p_i=\{q_i,a_{{\rm orb}, i},b_i,\theta_i\}$ 
that could have
given rise to it. We would also like to know, with some level of
confidence, if the observed \lc\ is not due to \bi\ lensing.

Therefore, 
given the \lc\ and its \ex , we proceed to identify all \bl\ \lc s that are
essentially indistinguishable from it where our criterion for
indistinguishability is set by Eqn.(2b), with $\epsilon$ determined by the
photometric error. For a given level of photometric uncertainty, $\delta$,
$\epsilon$ should be chosen to be roughly equal to $2\, \delta$,
since $\delta \times \{ \int_{-1}^1 dt\} \sim \epsilon$.

The simplest version of our method consists of
looking for \lc s that provide good fits to the original by 
randomly generating a large
number of lensing \ev s, $p_i$, 
finding the \lc\ expansion of each, and 
computing the distance, in the
space of the coefficients,  between the \lc\ and the original.
Those sets of physical parameters with \lc s which lie close to
the \lc\ of the original in the space of the coefficients,
are  candidates for the physical lensing \ev\ that gave rise
to the original \lc . 

 
\subsubsection{Improved Sampling}

A completely random sampling is numerically 
inefficient; it is preferable to first identify all regions of the 
parameter space in which there is some possibility of finding
a match. We can do this by maintaining a ``library'' of \ev\
characteristics.  
The library relates the physical parameters describing an \ev\ to the 
\lc\ parameters that characterize the resulting \lc .  
Each line of the library is specified by the 
values $\{q_i,a_{{\rm orb}, i},b_i,\theta_i\}$ used to generate a \lc ; the
other entries include the locations and values of the magnification at extrema
and inflection points.
We begin by searching
through the library for those \lc s whose gross features
are consistent with those of the observed \lc . 
Specifically,  we identify physical events 
(i.e., sets of physical parameters) 
that generate light curves with positions of extrema and inflection
points, and magnifications at these points, within roughly $10\%$
of those of the observed \lc. 
The physical parameters of 
each such library
member represent an event whose gross morphological
properties are similar to those of the observed \lc .
These points in the parameter space   
therefore provide a starting point for a finer sampling of a small  
region of the parameter space. It is this finer sampling which yields the
matching \lc s. Figure 2 shows three examples; 
the sampling is described in the 
caption. The degree to which the parameter space sampling can influence the 
discovery of degenerate solutions, and to which improved photometric
precision can help to break degeneracies, is demonstrated in Figures 2 and 3,
respectively. 

How many \ev s need to be included in the library, if we are to be certain 
that we have carried out a smooth sampling of the parameter space?
If the morphology of a \lc\ (i.e., positions of extrema, for example)
changed dramatically when the physical parameters specifying the \ev\
changed but little, a comprehensive library would need
to be large. Fortunately, the morphological features change in
a way that is gradual and consistent as the physical parameters
are changed (see Fig.4), so that a library with ${\cal O}(10^5)$ \ev s
appears to be adequate.   
\subsection{Singular or Sharply Peaked Light Curves} 

\vspace{-.2 cm}

Consider a case in which the
track of the source does not pass close to the caustic structure associated  
with the \bl . The \lc\ will be smooth and, generally,
only few terms are needed in order for the Tcheychev \ex\ to
provide a good approximation to it. (For example, $a_{10}$, the 
sixth even \ct\ in the \ex\ of the first \lc\ in
Figure 2 is only $1\%$ of $a_0$; the odd \ct s are all below the $1\%$
level.) If we allow the track of the source
to come closer to the caustic structure 
we will obtain \lc s that become ever
more sharply peaked. These \lc s can formally be expanded,
but an \ex\ to 
reasonable order will not provide a good approximation to the \lc .
(See the middle set of panels of Figure 2.)     
Nevertheless, the \ex\ \ct s, augmented by the the values of the 
gross morphological parameters, can indeed be used to characterize 
even the most extremely perturbed \lc .

Just as in the case of \lc s
observed to be smooth, we search through the library to identify
all sets of physical parameters $p_i$'s giving rise to \lc s with similar
gross features. If the gross features match to within 
a chosen level of precision ($\sim 10\%$) we then compute,
in the region around each such $p_i$, \lc s for large numbers of randomly
generated sets of physical parameters.
 When condition (2b) is satisfied,
we are guaranteed that the 
two \lc s are equivalent, to the level of precision we have chosen.

It is interesting to note that 
neither \lc\ will match the expansion, which has inflection points and
extrema not shared by the \lc s. But the two light curves and the \ex\
agree at the $m$ zeros of $T_m(t)$, and, in addition, the two \lc s agree
at all extrema and inflection points. The two light curves are
therefore equivalent. In general, 
we find that the shape of even extreme \lc s
is determined by the value of the \mag\ at a relatively small number of
points.      

\section{False Positives: Eliminating the Pretenders} 

\vspace{-.2 cm}

Given the tremendous diversity of \bi-\ml\ \lc s,  
it has been conjectured that
essentially {\it any} \lc\ can be well-fit by a \bi-lens model. 
If this conjecture is true, then it poses a serious problem for 
the analysis of data from the monitoring programs. The true rate
of lensing would never be known, since virtually any form of
stellar variability that is neither continuous nor 
repeating could be interpreted as
a microlensing event. Fortunately, the validity of this conjecture 
can be constrained in a quantifiable way. First we consider individual
\lc s, and then ensembles of \lc s.  

The morphology of a \lc\ is determined by the topology of the isomagnification
contours. This is determined by the physics of the \ml\ process.
Although many variations are associated with the many possible sets of
values of the physical parameters of the lensing events, it is not
true that a binary-lens light curve can take on any 
arbitrary functional form. 
There are a few obvious constraints. For example, 
although \lc s can be singular (or, taking finite-source-size effects into
account, nearly singular, exhibiting sharp rises and declines), the 
number of such singularities is constrained by the fact that each
represents a crossing through a closed caustic curve. Thus, there are always
an even number of caustic crossings. Even when the track of the \lc\
grazes the caustic structure, so that the time between the passage into
and out of the closed curve is small--perhaps too small to resolve--the 
magnification before and after the double passage
must be consistent with a grazing incidence, and this itself provides
constraints. Furthermore, the magnification within the caustic structure has 
a well-defined minimum (Witt \& Mao 1995). Similarly, the positioning and 
magnification of 
smooth peaks cannot be arbitrarily arranged. Even a function with
a single peak can be distinctive enough that a binary lens model
simply cannot fit. We have carried out a test to determine whether a 
specific blended point-lens \lc\ could be mis-identified as a
\bl\  \lc. First we generated a point-mass lens \lc;   
out of $10^5$ randomly generated binary-lensing events, we found that 753   
 were equivalent to it if the uncertainty in the photometry were
at the 2\% level. However, under the same conditions, no 
equivalent binary-lensing event was found for the
blended version of this light curve.
(We assumed that the lensed source contributed 10\% of the baseline flux.) 
This test provides a simple, 
quantifiable counterexample to the conjecture that any \lc\ can be fit by some
\bl\ model.

As in the example cited above, 
the failure to find good matches is operational, subject to 
the conditions under which the search has been conducted, which
must always be explicitly stated. 
With this caveat, we note that 
the mathematical formalism we have used provides a concrete
and well-defined way to test 
the conjecture that any specific function can be well-fit by a binary-lens
model. The level at which a search rules out a binary fit can be specified
and, if a \bl\ fit does work, the degeneracy of the physical solutions can
also be fully worked out. 

Another way to see that \bl\ \lc s do have distinctive
characteristics that cannot be mimicked by arbitrary functional forms, is to
examine the distribution of \bl\ \lc s in the parameter space defined 
by the coefficients $\{a_n\}$. The conjecture that any \lc\ can be fit 
by a \bl\ model  
is equivalent to the conjecture that the \ct\ space is everywhere dense
in points corresponding to \bl\ \lc s. In Figure 5 we study the pattern
of points in the coefficient space associated with smooth double-peak
\lc s. 
It is obvious that there is a pattern, and not a uniform distribution.
The pattern is so distinctive that it is possible to clearly 
detect the presence of a cloud of points that is somewhat deviant from it.
This cloud of points actually defines a distinct
distribution: double-peak \lc s in which the magnification dips below
$1.34$ between peaks, so that one is in essence watching two separate
lensing events
 (\rd\ \& Mao 1996). Thus, the figure demonstrates   
that the \ct\ space distribution of points associated with
\bl\ \lc s contains useful information. In addition, Figure 5
shows that blending serves to shift the distribution of
\lc s in the parameter space. This is another demonstration of the link
between the physical characteristics of a \lc, and the position
of the corresponding point in the space of the coefficients.

\section{Applications and Extensions}
\vspace{-.2 cm}

\subsection{Applications}
\vspace{-.2 cm}

What are the prospects that \bl\ \ev s that might otherwise have been
missed, will be successfully identified via our method? 
It is already known that the fit between the data and lens model for
some \lc s 
identified as being due to lensing by a single mass, can be
marginally improved by using a \bl\ model (see, e.g., Dominik \& 
Hirshfeld 1996).  
In such cases, it is questionable whether the introduction of
new parameters to effect a marginal improvement should be
viewed as strongly supporting the hypothesis that the lens
was actually a binary. 
The key to a successful model is that the fit to the data should be
convincingly better than to a point-lens model.      
The \lc\ must therefore be distinctive (as the OGLE 7 event was, for
example, and as many smooth \bl\ \lc s are), or the photometry 
and sampling must be 
better than is typical for the \mo\ teams.  
Since we cannot easily order up distinctive \ev s, we can act to improve
the situation only through improved monitoring. Fortunately, 
the benefits of more frequent sampling of ongoing \ev s and real-time
analysis are now being realized through the combined efforts of the
\mo\ teams, who announce their discovery of \ev s as soon as
possible, and independent groups of observers who have designed 
observational \pr s to provide frequent 
coverage of ongoing events (see, e.g., Albrow {\it et al.} 1996; 
Alcock {\it et al.} 1996). In principle, the follow-up teams, who
have access to telescopes positioned around the Southern hemisphere,
can monitor \ev s hourly, from the time of first notification until the
\ev\ is finished. Furthermore, because the follow-up teams are
most interested in deviant \lc s, particularly if there is a chance that
they are due to lensing by a planetary system, their photometric    
measurements tend to be sensitive, with uncertainties at the $1-2 \%$ level.
During their 1996 observing season, the PLANET collaboration
followed the progress of 2 events in the LMC and 31 in the direction of the
Galactic Bulge; in 1995 they studied 41 Bulge \ev s 
(www.astro.rug.nl/\~\ planet).  
The ascendancy of these follow-up \pr s and proposals to implement
even more sophisticated \mo\ \pr s,  improve our chances of detecting
deviant \lc s. 
The next step in our research program is therefore to 
work with several of the teams to fashion an application of the method
which carefully considers and meshes the issues of observational and
\ct\ space sampling.

\subsection{Other Perturbations}
\vspace{-.2 cm}

The method we have developed can be applied to \lc s that differ from the
standard because of any astronomical effect, e.g., blending, parallax,
finite source size. The more complicated the effect (such as finite source
size) or the more effects we want to consider, the larger the number
of relevant physical parameters becomes. Let's therefore augment the
set of physical parameters, $\{p_i\}$, by adding an additional set of
parameters, $\{P_i\}$ to describe all of the other physical
effects that may be relevant. 
Since the method outlined in \S 2 is completely general, an obvious way to
proceed would be to simply use a parallel approach, including the full
set of parameters, $\{p_i,P_i\}$, instead of just $\{p_i\}$. 
This brute force approach would generally not be computationally
efficient, however.
The library of events for the preliminary comparison would be large,
and the regions we would need to subsequently sample in the parameter
space would be of high dimension. 

Especially when the additional effects can be easily expressed analytically,
as they can, e.g., for blending and for velocity effects 
(including parallax and the
orbital motion of the lens if it is a binary) there is a more efficient
way to proceed.       
Because the influence of these additional effects on the \lc\ 
can be expressed analytically, we could, if we knew the values of the $\{P_i\}$,
effectively eliminate them, to derive a residual 
\lc\ having either the form of a point-lens \lc\ or of a multiple
system, such as a \bl . 
Of course we don't know the values of the $\{P_i\}$ {\it a priori}.
But the form of the observed \lc\ generally allows us to set some
limits on the range of values of each parameter that might be relevant.
Thus for each of a randomly sampled set of parameters, $\{P_i\}$,
within the appropriate
range, we can derive a ``residual" \lc\ and then apply the 
method outlined in \S 2. 
    
\subsection{Conclusion}
\vspace{-.2 cm}

The principle features of an approach to systematically study
\lc s perturbed because of lens binarity or any other astronomical
effect have been outlined. The application of the 
approach has been demonstrated.

One important feature of this work is that it provides a unified
approach for the identification and study of all \bl\ \lc s,
and indeed, all \lc s perturbed from the standard point-mass lens
form. 
Perhaps equally important, is that it provides a systematic way to
quantify the degeneracy of the physical situations which
could possibly have given rise to any particular
observed \lc . 
This is important because, unless the extent of the  degeneracy is fully 
understood,
it is not possible to use an ensemble of \lc s
to place meaningful constraints on the true distribution of lens
parameters.    
We have found that, as may well be expected,
the level
of degeneracy for \bl\ \lc s is highest for \lc s with the
fewest distinctive features, i.e., those most like the 
standard point-mass lens \lc s. This result emphasizes the
challenge of distinguishing point-lens \ev s from \bl\ \ev s,
and thereby also serves to reinforce the need for the 
type of frequent, sensitive follow-up which has been tested
during the past two years (see, e.g., Albrow {\it et al.} 1996;
Alcock {\it et al.} 1996). We note, in addition, that a large
fraction of \bl\ \lc s are neither close to the point-lens
form, nor as extremely perturbed as the \bl\ \lc s
discovered in the data so far. These are the \lc s whose
apparent absence in the data set is puzzling.
It remains to be determined if the lack of such \ev s may 
be indicative of problems with the detection efficiency, 
perhaps related to the use of a set of
detection criteria that is too narrow.

The challenge for the future is to use the principles
outlined here to 
study real data. There are almost certainly
\lc s in the present data set which could tell us more about the 
\ev s that gave rise to them through application of a method such
as the one presented here. There will certainly be \lc s observed
by the follow-up teams for which this will be the case.
Applying these principles to the data will guide us to more refined
and efficient approaches.
But, most importantly, work along these lines will help us to
learn more about the population of stellar and dark lenses in the Local
Group.
   
\bigskip

\centerline{\bf Acknowledgments} 

This work was supported in part by NSF GER-9450087. We also thank the anonymous
referee for comments that helped us to clarify the presentation.

\vfil\eject
\noindent{Figure 1.--\
These $15$ \lc s have been randomly
generated, with all of the physical variables (\S 2.1)
chosen from
uniform probability distributions.
The mass ratio was chosen to have
values ranging from $0$ to $1$, the orbital separation varied between
$0$ and $1.8 R_E$, the inclination of the source track relative to the binary
axis varied from $0$ to $2\,\pi$, and the distance of closest
approach varied between $0$ and $1 R_E$.
The only restriction placed on the \lc s shown here, is that
the integrated area under the \lc\ must differ by more than $1\%$
from the area under the point-mass lens light curve most similar
to the one shown. By the point-lens light curve most similar to the one shown,
we mean the point-lens light curve that matches exactly in the wings
(i.e., at $x=+1$ and $x=-1$) and that bounds the same area from above.
}

\noindent{Figure 2.--\ {Three \bl\ \lc s (left set of panels)
were generated by three physical events,
specified by the values $\{q_0,a_{{\rm orb},0},b_0,\theta_0\}$. 
A Monte Carlo approach
was used to sample the entire parameter space of physical variables
to find other \ev s that generate nearly identical \lc s. Each of these
new \ev s served as a seed for further sampling of a small region. 
The middle set of panels show all
the physical \bis\ ($a_{\rm orb}$ vs $log(q)$) we found that can, with
appropriate choices of $b$ and $\theta$, give rise to \lc s whose 
integrated deviation from the one shown is smaller than $2\%$.  
Note that the level of degeneracy generally declines as the number of
distinctive \lc\ features (extrema and inflection points) increases.
Other points of interest: (1) good uniform sampling is needed in the  
first step of the search to prevent the clumpiness produced by our seeding
in the top middle panel. (2) Binary separations of $2 R_E$ still yield
\ev s with clear evidence of lens binarity.
Finally, the set of panels on the right show 
the first 50 Tchebyshev coefficients
for the corresponding light curve in the left-most panel. The coefficients are
expressed in units of $a_0$. Note that the smoother the light curve,
the faster the convergence.
}

\noindent{Figure 3.--\  
The integrated deviation between the top \lc\ in Figure 1 and the ``nearest"
point-mass \lc\ is $\sim 1.5 \%$. This is shown in the top panel, with the 
point-by-point difference between the \lc s plotted in the middle panel.
The bottom panel illustrates that, even if the photometry were good enough
to distinguish between the \bl\ and point-mass \lc s
shown in the top panel, the \bl\ \lc\
would still be degenerate. Points in the bottom panel are associated
with \lc s that differ from the original by less than $1 \%$.
}

\noindent{Figure 4.--\ 
Variation in the amplitude of a light curve as the parameters are varied.
In the left panel the parameters $a_{\rm orb}, \theta, q$ are kept fixed
while the impact parameter $b$ changes by 0.1 from light curve to \lc.
In the set of panels on the right, the parameters $a_{\rm orb}, q, b$ are kept fixed
while $\theta$ changes by $\pi/6$ from light curve to \lc .
}

\noindent{Figure 5.--\ 
Points in the \ct\ space of all smooth double-peak 
\bl\ \lc s found in a sampling of
$\sim 200,000$ \ev s with $b<1 R_E$. $a_0$ has been normalized to unity.  
The top left panel is $a_{4}$ vs $a_2$; in the second row are $a_{6}$ vs $a_2$,
and $a_{6}$ vs $a_4$, respectively; the ordering of axes continues
in this way in rows 3 and 4.    
Note that, not only is the general pattern of points striking,
but also that deviations from it can be specifically
linked to \lc\ features. The diffuse cluster of points  near the origin
in each panel corresponds to smooth double-peak \lc s that would likely
be perceived as repeating events. The main pattern of points traces the placement of 
continuous smooth perturbations with $A>1.34$ throughout. This
is demonstrated in the top two panels on the right. The panel just below these
shows that, even in a single plot, the effects of blending can be
distinguished. The dots are the same as in the top left panel; the 
circles are blended, smooth, 2-peak \bl\ \ev s. Peak placement and relative peak
and valley \mag s are the same as for the other \bl\ \lc s, thus representing
a particularly mild perturbation of standard \bl\ \lc s; it is 
therefore interesting that the difference can be distinguished on this scale.
}
\vfil\eject
\begin{figure}
\plotone{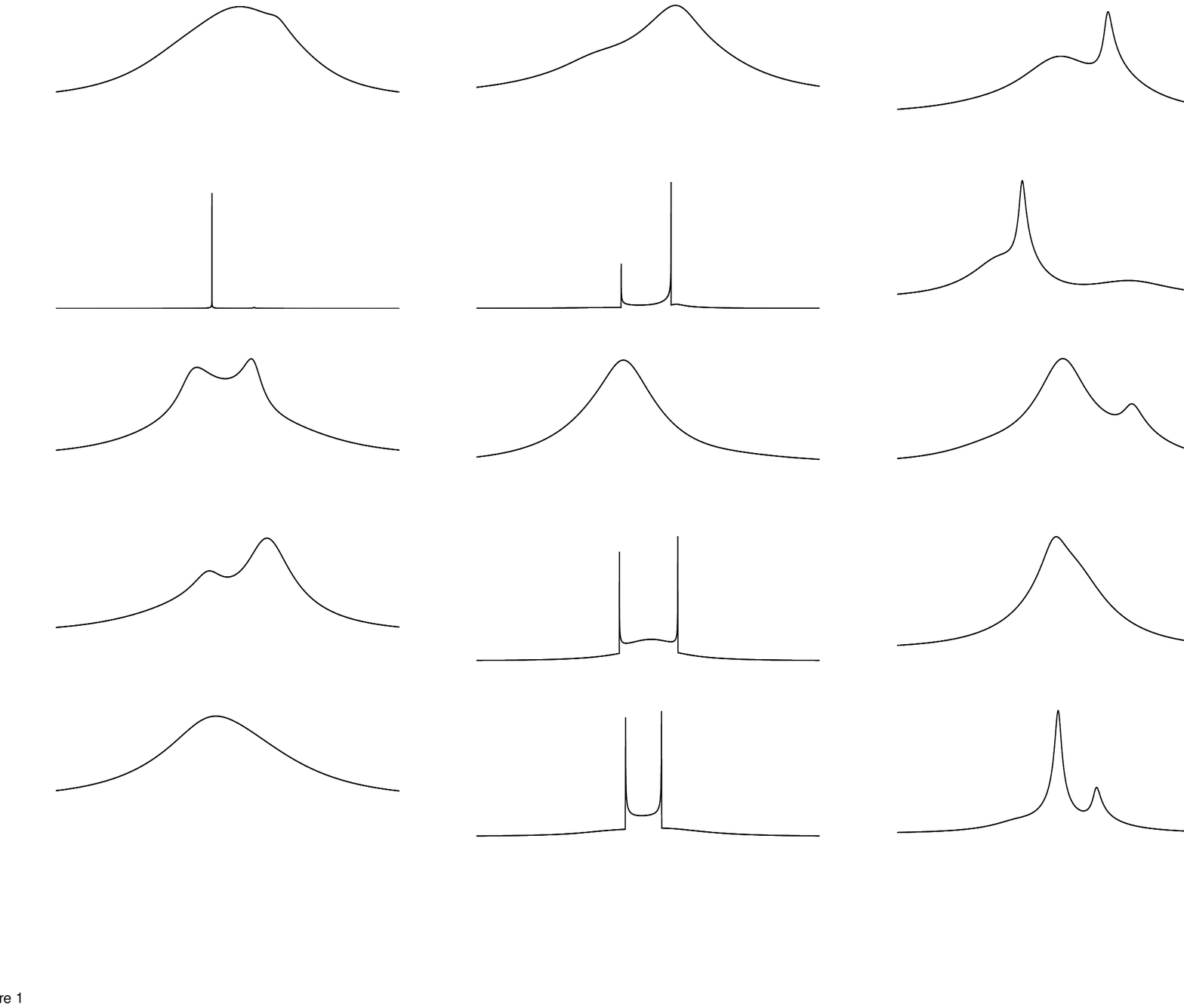}
\end{figure}
\vfil\eject 
\begin{figure}
\plotone{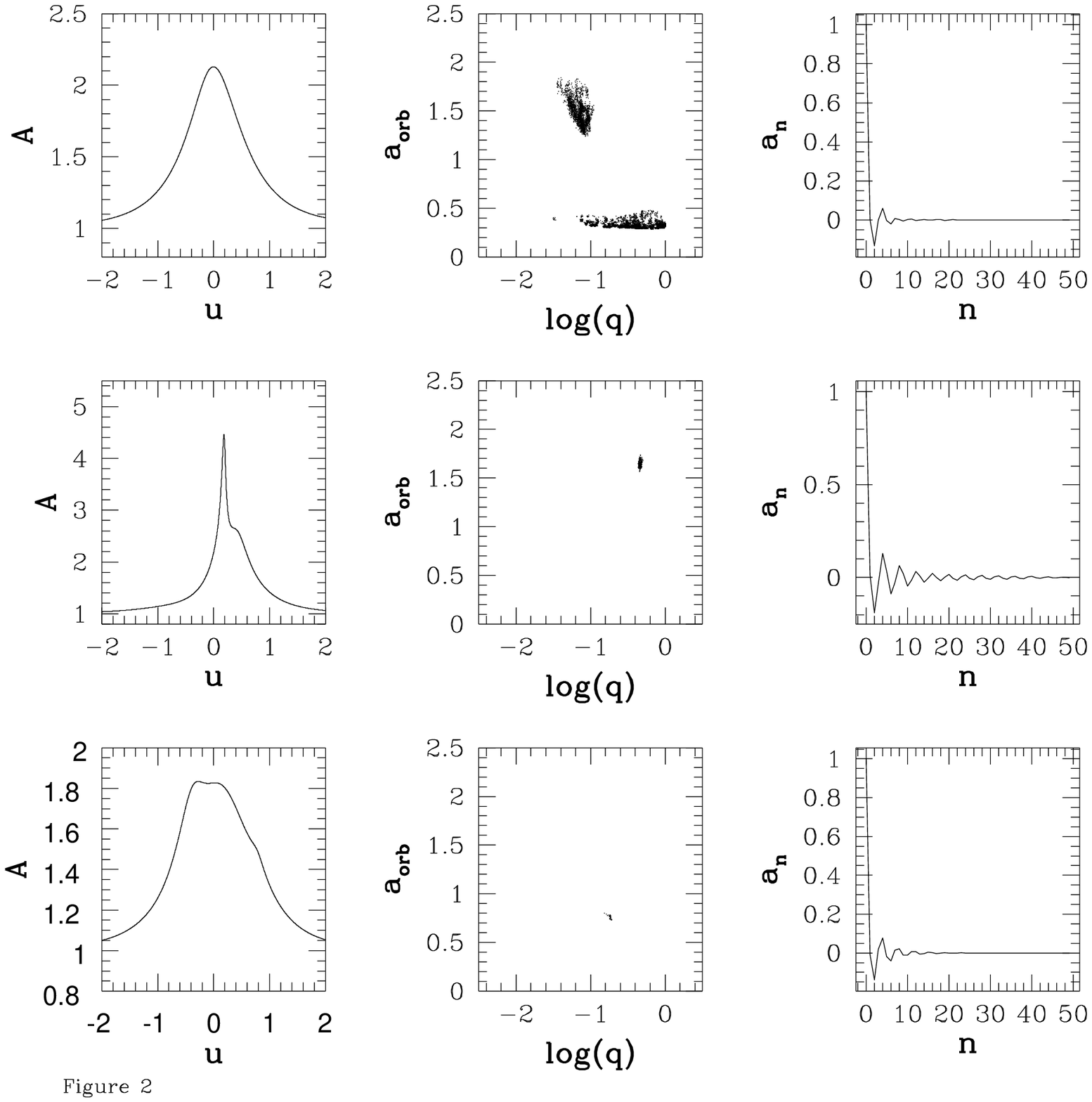} 
\end{figure}
\vfil\eject  
\begin{figure}
\plotone{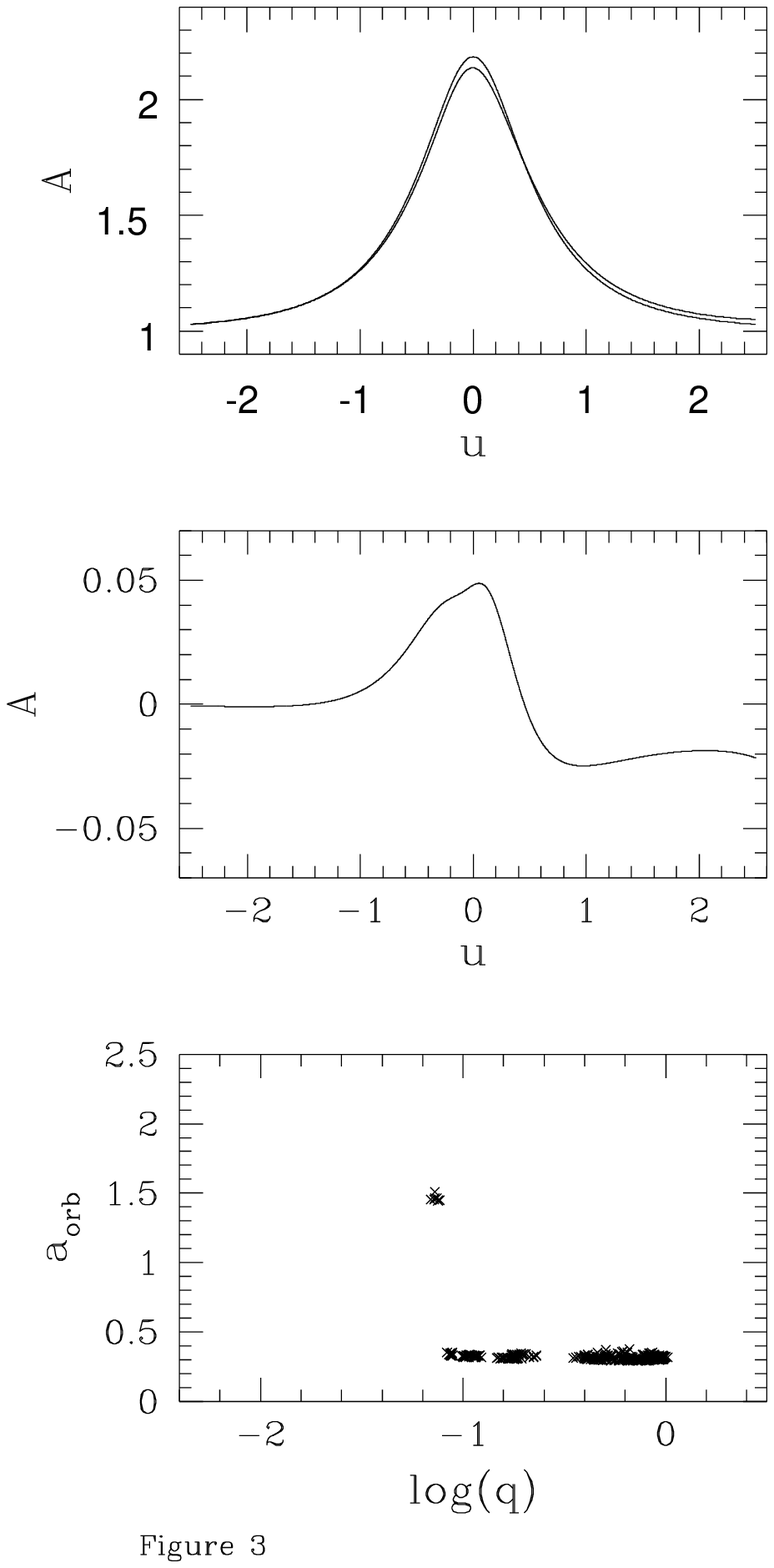} 
\end{figure} 
\vfil\eject 
\begin{figure}
\plotone{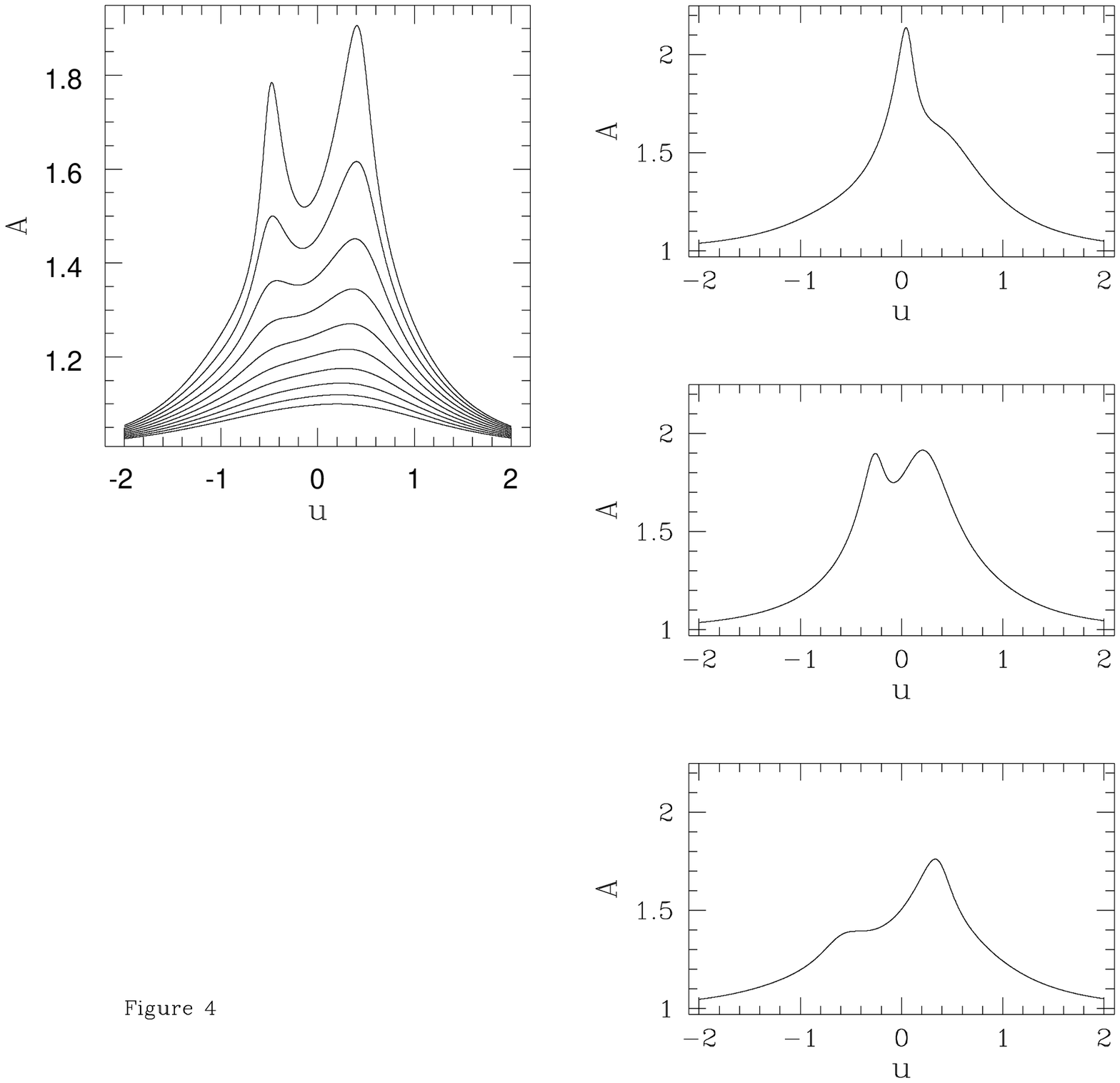} 
\end{figure} 
\vfil\eject 
\begin{figure}
\plotone{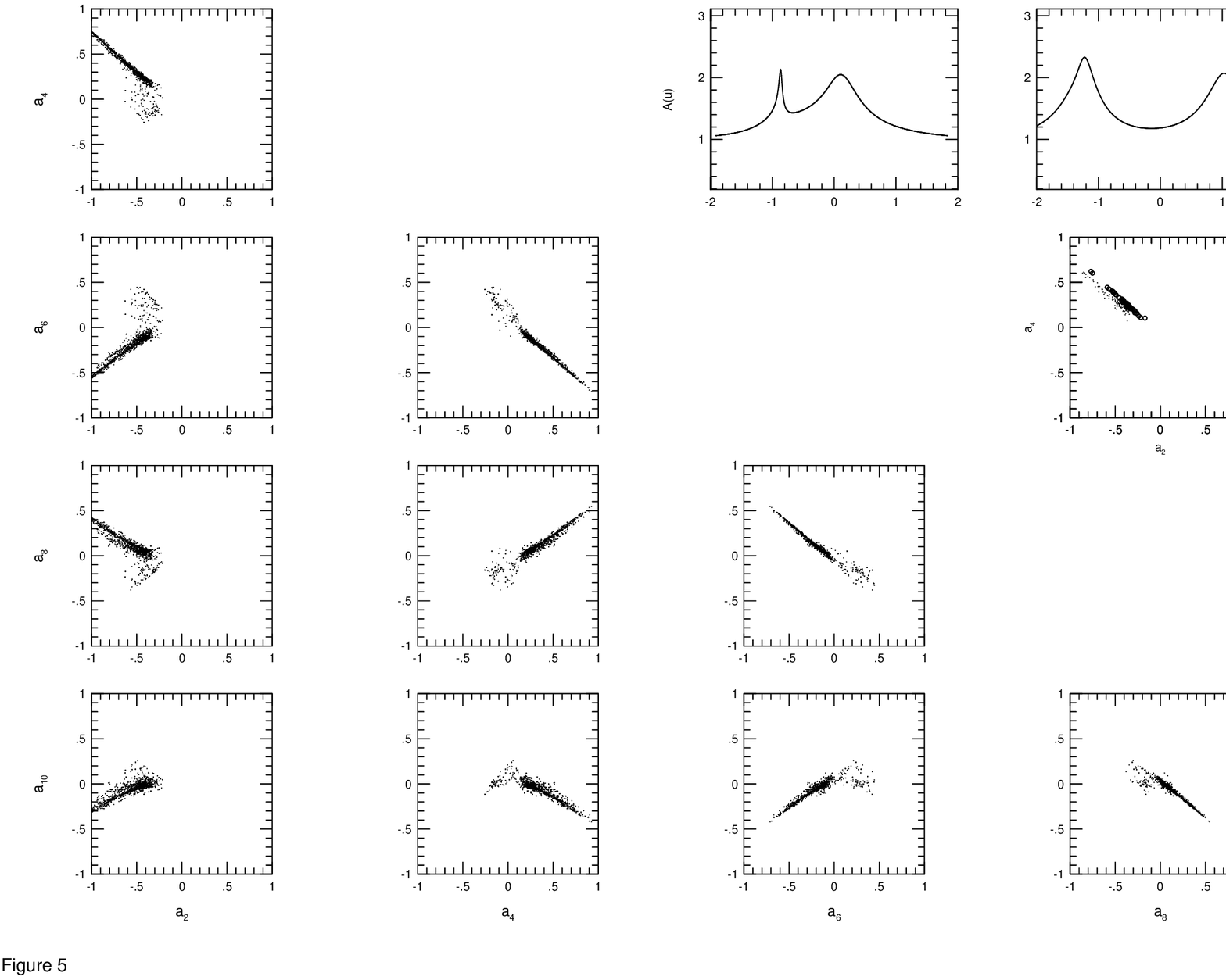} 
\end{figure}

\end{document}